\pdfoutput=1

\documentclass[preprint,5p,numbers,sort&compress]{elsarticle}

\usepackage{graphicx}
\usepackage{amssymb,amsmath}
\usepackage{natbib}
\usepackage{hyperref}
\usepackage{multirow}
\hypersetup{pdftitle = The title of my PDF, pdfauthor = My name, pdfsubject= The subject, pdfkeywords = keyword1 keyword2 keyword3}
\hypersetup{colorlinks = true, linkcolor = blue, anchorcolor = red, citecolor = blue, filecolor = red, pagecolor = red, urlcolor = blue}

\begin{document}

\begin{frontmatter}
\title{Equilibrium points and basins of convergence in the triangular restricted four-body problem with a radiating body}

\author[gag]{J. E. Osorio-Vargas}

\author[gag]{Guillermo A. Gonz\'alez}

\author[fld]{F. L. Dubeibe \corref{cor1}}
\ead{fldubeibem@unal.edu.co}

\cortext[cor1]{Corresponding author}

\address[gag]{Grupo de Investigaci\'on en Relatividad y Gravitaci\'on, Escuela de F\'isica, Universidad Industrial de Santander, A.A. 678, Bucaramanga 680002, Colombia}

\address[fld]{Grupo de Investigaci\'on Cavendish, Facultad de Ciencias Humanas y de la Educaci\'on, Universidad de los Llanos, Villavicencio 500017, Colombia}

\begin{abstract}

%In the current paper, we aim to extend the work of Baltagiannis [12] and Zotos [13] by performing a full anal- ysis of the location, stability, and basins of convergence of the equilibrium points, associated to the inclusion of a radiating body in the equilateral triangle configuration of the four-body problem. The inclusion of the radiation and drag forces in the restricted four-body problem, allow us to model in a more realistic way the dynamics of a test particle in presence of an astrophysical system with, e.g., an active star. Unlike the gravitational force, radiation and drag forces are generally non-conservative, causing a loss of orbital energy such that the particles will spiral to- ward the source. This new effect can significantly modify the dynamics of the model and deserves a complete study in the context of the restricted four-body problem.

The dynamics of the four-body problem have attracted increasing attention in recent years. In this paper, we extend the basic equilateral four-body problem by introducing the effect of radiation pressure, Poynting-Robertson drag, and solar wind drag. In our setup, three primaries lay at the vertices of an equilateral triangle and move in circular orbits around their common center of mass. Here, one of the primaries is a radiating body and the fourth body (whose mass is negligible) does not affect the motion of the primaries. We show that the existence and the number of equilibrium points of the problem depend on the mass parameters and radiation factor. Consequently, the allowed regions of motion, the regions of the basins of convergence for the equilibrium points, and the basin entropy will also depend on these parameters. The present dynamical model is analyzed for three combinations of mass for the primaries: equal masses, two equal masses, different masses. As the main results, we find that in all cases the libration points are unstable if the radiation factor is larger than 0.01 and hence able to destroy the stability of the libration points in the restricted four-body problem composed by Sun, Jupiter, Trojan asteroid and a test (dust) particle. Also, we conclude that the number of fixed points decreases with the increase of the radiation factor. 

\end{abstract}


\end{frontmatter}

\section{Introduction}
\label{intro}

The $n$-body problem is one of the oldest and most researched problems in astrophysics, it deals with the motion of bodies that are subject to their mutual gravitational attractions (or to other forces as well). The study of the $n$-body problem is essential to the understanding of the motions of celestial bodies and is of particular interest in spacecraft navigation \cite{Marchand2007}. Since there is no general analytical solution to the $n$-body problem for $n\ge 3$, several simplifications have been introduced with the most prominent being the restricted three and four body problems \cite{Meyer2008}. In both cases, the mass of one of the bodies (test particle) is negligible in comparison to the others (primaries) such that it does not affect their motion. In the restricted three-body problem the primaries move in a circular or elliptic trajectory (solutions of a two-body problem), while in the restricted four-body problem the primaries move in a straight-line equilibrium configuration or in an equilateral triangle configuration (solutions to the three-body problem) \cite{Micha1981}.

The restricted four-body problem has many applications in celestial mechanics, dynamical astronomy, and galactic dynamics, mainly motivated by the fact that there are many astrophysical systems that can be roughly approximated to that model, for example, the Sun-Jupiter-Saturn-satellite system, the Sun-Jupiter-Trojan-spacecraft system, the Sun-Jupiter-Earth-satellite system, the Saturn-Tethys-Telesto-spacecraft system \cite{Alvarez2015}, or even any of the four-star systems of our galaxy \cite{Roy2004}. Aiming to get a better understanding of the problem, several modifications to the effective potential in the classical restricted four-body problem have been proposed, see {\it e.g.} \cite{Kalvouridis2007, Papadouris2013, Kumari2013, Asique2015}, where it was found that the introduction of the additional terms significantly modifies the existence, position, and stability of the equilibrium points, and therefore the overall dynamics of the system. 

Since it is a well-known fact that in the $n$-body with $n\ge 3$ it is not possible to find explicit formulae for the position of the equilibrium points, numerical methods become the natural and the most efficient way of finding the locations of the fixed points in many different dynamical systems. One of the most popular methods to find roots of multivariate functions is the Newton-Raphson algorithm, this method allows us to get a global picture of the set of initial conditions that lead to a particular fixed point. The final picture produced after evolving (via the root-finding algorithm) the whole set of initial conditions in a certain region $\Omega$ of the configuration space is called the basin of convergence (see {\it e.g.} \cite{Zotos2017a, Zotos2018}). 

During the past few years, the Newton-Raphson basins of convergence have been extensively investigated in many different versions of the restricted four-body problem. As characteristic examples of these works, may be mentioned the one by Baltagiannis \& Papadakis \cite{Baltagiannis2011}, who determined the position and stability of the equilibrium points in the equilateral triangle configuration of the four-body problem using different combinations of primaries, and the study carried out by Zotos \cite{Zotos2017b}, who extended the study of the basins of convergence for cases not considered by Bataglianis. Moreover, Suraj {\it et. al.} \cite{Suraj2017a} studied the existence and locations of libration points along with the Newton-Raphson basins of convergence for the same model, but using triaxial rigid bodies as primaries, and later studied the effect of small perturbations in the inertial forces on the Newton-Raphson basins of convergence \cite{Suraj2017b}. Further, the linear equilibrium configuration of the restricted four-body problem was considered by Zotos in order to determine how the mass parameter and angular velocity influence the geometry of the Newton-Raphson basins of convergence \cite{Zotos2017c}, while the photo-gravitational restricted four-body problem with variable mass was studied in \cite{Mittal2018}.

In the current paper, we aim to extend the work of Baltagiannis \cite{Baltagiannis2011}  and Zotos \cite{Zotos2017b} by performing a full analysis of the location, stability, and basins of convergence of the equilibrium points, associated to the inclusion of a radiating body in the equilateral triangle configuration of the four-body problem. The inclusion of the radiation and drag forces in the restricted four-body problem, allow us to model in a more realistic way the dynamics of a test particle in presence of an astrophysical system with, {\it e.g.}, an active star. Unlike the gravitational force, radiation and drag forces are generally non-conservative, causing a loss of orbital energy such that the particles will spiral toward the source. This new effect can significantly modify the dynamics of the model and deserves a complete study in the context of the restricted four-body problem.

The present paper has the following structure: the equations of motion for a test particle are derived in section \ref{sec2}. In section \ref{sec3}, we analyze the position, linear stability, zero velocity surfaces and basins of convergence of the equilibrium points, as a function of the radiation parameter. Here, we study three combinations of mass for the primary bodies: equal masses, two equal masses, different masses. In Section \ref{sec4} the complexity (unpredictability) of the basins is analyzed through the basin entropy, showing its dependence with the radiation parameter. Finally, the main conclusions of this work are drawn in section \ref{sec5}.

%%%%%%%%%%%%%%%%%%%%%%%%%%%%%%%%%%%%%%%%%%%%%%
\section{Equations of motion}\label{sec2}
%%%%%%%%%%%%%%%%%%%%%%%%%%%%%%%%%%%%%%%%%%%%%%

Consider the equilateral restricted four-body problem (henceforth ERFBP). Let $m_1, m_2$, and $m_3$ be the masses of the primaries and let $m$ be the mass of the test particle, which satisfies $m\ll m_{i}$ with $i=1,2,3$. The primaries revolve in the same plane with uniform angular velocity, and regardless of the mass distribution of the primaries, they always will lie at the vertices of an equilateral triangle. If the primary $m_1$ is a radiating body, the equations of motion for the test particle $m$ in an inertial frame of reference $\bold{R}=(X, Y)$, can be written as \cite{Burns1979}:
\begin{eqnarray}
m\,\ddot{\bold{R}} &=& -\sum_{i=1}^3 \frac{G m m_i}{R_i^3} \, \bold{R}_i + \frac{S A Q_{\text{pr}}}{c} \Bigg[ \frac{\bold{R}_1}{R_1} \nonumber \\
&-& (1 + sw)  \left( \frac{\dot{\bold{R}}_1 \cdot \bold{R}_1}{c \, R_1} \dfrac{\bold{R}_1}{R_1} - \frac{\dot{\bold{R}}_1}{c}\right) \Bigg],
\label{eq:motion_gen}
\end{eqnarray}
where $S$ denotes the solar energy flux density, $A$  the geometric cross-section of the test particle, $Q_{\text{pr}}$ the radiation pressure coefficient, $c$ the speed of light, and $sw$ the ratio of solar wind to Poynting-Robertson drag.

In Eq. (\ref{eq:motion_gen}), the first term on the right-hand side represents the influence of the gravitational forces due to the three massive bodies, while the second term includes the effect of radiation pressure and drag forces. The ratio of force due to radiation pressure compared to gravity, allow us to introduce a dimensionless radiation factor (see {\it e.g.} \cite{Kumari2013})
\begin{equation}
\beta = \frac{F_r}{F_g} = \frac{S A Q_{\text{pr}} R_1^2}{c \, G M m_1},
\label{eq:beta}		
\end{equation}

For simplicity, in all that follows we shall use canonical units, such that the sum of the masses, as well as the distance between the primaries, the angular velocity, and the gravitational constant, are set to 1. Additionally, as we consider the non-relativistic limit of the model, the speed of light will be chosen to the value $c = 1\times 10^{4}$ \cite{Dubeibe2017,Dubeibe2018}, unless otherwise is specified. Taking into account the previous definitions, the equations of motion in a synodic frame of reference $\bold{r}=(x,y)$ read as
\begin{eqnarray}
\nonumber \ddot{x} &-& 2 \, \dot{y} = x - \dfrac{m_1 \, (x - x_1)}{r_1^3} \, (1 - \beta) - \dfrac{m_2 \, (x - x_2)}{r_2^3}\\
&-& \dfrac{m_3 \, (x - x_3)}{r_3^3} - (1 + sw) \, F_x \, ,\label{eq:motion_1}\\
\nonumber \ddot{y} &+& 2 \, \dot{x} = y - \dfrac{m_1 \, (y - y_1)}{r_1^3} \, (1 - \beta) - \dfrac{m_2 \, (y - y_2)}{r_2^3}\\
&-& \dfrac{m_3 \, (y - y_3)}{r_3^3} - (1 + sw) \, F_y \, ,
\label{eq:motion_2}
\end{eqnarray}
where
\begin{eqnarray}
\nonumber F_x &=& \dfrac{\beta \, m_1}{c \, r_1^2} \, \left[ \dfrac{(x - x_1)}{r_1^2} \, N + \dot{x} - (y - y_1) \right],\\
\nonumber F_y &=& \dfrac{\beta \, m_1}{c \, r_1^2} \, \left[ \dfrac{(y - y_1)}{r_1^2} \, N + \dot{y} + (x - x_1) \right],
\end{eqnarray}
with 
\begin{equation*}
N = (x - x_1) \, \dot{x} + (y - y_1) \, \dot{y} \, ,
\end{equation*}
and 
\begin{equation*}
{r}_i = \left[ (x - x_i)^2 + (y - y_i)^2 \right] ^{1/2} ; \quad i = 1, 2, 3. 
\end{equation*}
 
The coordinates of the primaries can be fully determined by placing the center of mass of the ERFBP at the origin of the synodic frame of reference, with $m_1$ lying on the $x$-axis, and taking each side of the equilateral triangle equal to unity (see Eq. (13) in Ref. \cite{Moulton1900}), {\it i.e.},
\begin{eqnarray*}
y_1 &=& 0,\\
m_1 \, x_1 + m_2 \, x_2 + m_3 \, x_3 &=& 0,\\
m_1 \, y_1 + m_2 \, y_2 + m_3 \, y_3 &=& 0,\\
(x_2 - x_1)^2 + (y_2 - y_1)^2 &=& 1,\\
(x_3 - x_2)^2 + (y_3 - y_2)^2 &=& 1,\\
(x_1 - x_3)^2 + (y_1 - y_3)^2 &=& 1.
\end{eqnarray*}
This set of equations lead to the following solutions for the coordinates of the primaries
\begin{eqnarray}
x_1 &=& K_1 / K_2 \, ,\nonumber\\
x_2 &=& \dfrac{m_3 \, (m_2 - m_3) + m_1 \, (2 \, m_2 + m_3)}{2 \, K_1 \, K_2} \, ,\nonumber\\
x_3 &=& \dfrac{m_2 \, (m_3 - m_2) + m_1 \, (m_2 + 2 \, m_3)}{2 \, K_1 \, K_2} \, ,\nonumber\\
y_1 &=& 0 \, ,\nonumber\\
y_2 &=& \dfrac{\sqrt{3}}{2} \, \dfrac{m_3}{K_1} \, ,\nonumber\\
y_3 &=& \dfrac{\sqrt{3}}{2} \, \dfrac{m_2}{K_1} \, ,
\label{eq:coordinates}
\end{eqnarray}
with
\begin{equation*}
K_1 = \pm \sqrt{m_2^2 + m_2 \, m_3 + m_3^2} \,\,\, ; \,\, K_2 = m_1 + m_2 + m_3.
\end{equation*}
Therefore, there are four possible cases for the location of the three massive bodies. Here, we will consider the fourth case (see Table \ref{table1}).
\begin{table}[h]\centering
{\begin{tabular}{|ccccccc|}
\hline
{\textsc{Case}} & {$x_1$} & {$x_2$} & {$x_3$} & {$y_1$} & {$y_2$} & {$y_3$} \\ %\cellcolor[rgb]{0.33,0.41,0.58}\textcolor{white}
\hline\hline
1 & $-$ & + & + & 0 & $-$ & +\\\hline
2 & $-$ & + & + & 0 & + & $-$\\\hline
3 & + & $-$ & $-$ & 0 & $-$ & +\\\hline
4 & + & $-$ & $-$ & 0 & + & $-$\\\hline
\end{tabular}}
\caption{Location of the primary bodies according to the four possible cases. \label{table1}}
\end{table}

It can be easily noted that by defining the effective potential
\begin{equation}
U = \dfrac{1}{2} \, \left( x^2 + y^2 \right) + \dfrac{m_1}{r_1} \, (1 - \beta) + \dfrac{m_2}{r_2} + \dfrac{m_3}{r_3},
\end{equation}
the equations of motion (\ref{eq:motion_1}) and (\ref{eq:motion_2}) can be written in the compact form
\begin{eqnarray}
\ddot{x} - 2 \, \dot{y} &=& \dfrac{\partial U}{\partial x} - (1 + sw) \, F_x \, ,\label{eqm1}\\
\ddot{y} + 2 \, \dot{x} &=& \dfrac{\partial U}{\partial y} - (1 + sw) \, F_y \,.\label{eqm2}
\end{eqnarray}

Due to the existence of dissipative terms, the Jacobi constant is no longer conserved and varies with time according to the following relation
\begin{equation}
\dot{\cal C} =  2 (1 + sw) \left(F_x \dot{x}+F_{y} \dot{y}\right).
\label{eq:jacobi_variation}
\end{equation}

In absence of dissipative forces ($\beta = 0$), Eqs. (\ref{eqm1}) and (\ref{eqm2}) reduce to the classical ERFBP (see {\it e.g.} \cite{Moulton1900, Baltagiannis2011, Zotos2017b}), and hence the Jacobi constant (\ref{eq:jacobi_variation}) is conserved
\begin{equation}
{\cal C} = 2 \, U - \left(\dot{x}^2 + \dot{y}^2\right).
\label{eq:jacobi_c}
\end{equation}

%%%%%%%%%%%%%%%%%%%%%%%%%%%%%%%%%%%%%%%%%%%%%%
\section{Libration points and zero velocity surfaces}
\label{sec3}
%%%%%%%%%%%%%%%%%%%%%%%%%%%%%%%%%%%%%%%%%%%%%%

In what follows, we will determine how the radiation factor $\beta$ affects the position, stability, and basins of convergence of the libration points, when using one of the three possible combinations of mass for the primaries, $m_1=m_2=m_3, m_1\neq m_2=m_3,$ and $m_1\neq m_2 \neq m_3$. To do so, let us derive some general expressions in terms of the masses of the primary bodies. 

First, the location of the libration points can be determined by solving the system of equations $\dot{x} = \dot{y} = \ddot{x} = \ddot{y} = 0$, which leads to the following algebraic system of equations 
\begin{eqnarray}
&&x - \frac{m_1(x - x_1)}{{r}_1^3}  (1-\beta) + (1 + sw) \frac{\beta m_1 (y - y_1)}{c {r}_1^2} \nonumber\\
&&- \sum_{i=2}^3 \frac{m_i (x - x_i)}{{r}_i^3} = 0\,,\label{eq:equil_x}\\
&&y - \frac{m_1 (y - y_1)}{{r}_1^3} \, (1-\beta) - (1 + sw) \frac{\beta m_1 (x - x_1)}{c {r}_1^2}\nonumber\\
&&- \sum_{i=2}^3 \frac{m_i (y - y_i)}{{r}_i^3}  = 0\,,\label{eq:equil_y}
\end{eqnarray}
whose solution shall depend on the parameters $\beta$ and $s w$, as well as the values of mass, where the last ones will modify also the position of the primaries $(x_i, y_i)$ according to \eqref{eq:coordinates}. 

In spite of the fact that the Jacobi constant is not a conserved quantity when radiation terms are included ($\beta\neq 0$), an analytic expression for the zero velocity surfaces (ZVS) can be derived as follows. From Eq. \eqref{eq:jacobi_variation}, we get
\begin{eqnarray}
\nonumber C(t) &=& 2 U - (\dot{x}^2 + \dot{y}^2)\\
&-& 2 (1 + sw) \, \frac{\beta m_1}{c} \left[a_1 + a_2 + a_3\right],
\label{eq:jac_time}
\end{eqnarray}
with
\begin{eqnarray}
 a_1 &=& \int \dfrac{\left[ (x - x_1) \, \dot{x} + (y - y_1) \, \dot{y} \, \right]^2}{\text{r}_1^4} \, dt,\\
 a_2 &=& \int \dfrac{\dot{x}^2 + \dot{y}^2}{\text{r}_1^2} \, dt\\
 a_3 &=& \arctan \Bigg( \dfrac{y - y_1}{x - x_1} \Bigg),
\end{eqnarray}
and setting $\dot{x}=\dot{y}=0$, time dependent terms vanish and  consequently Eq. (\ref{eq:jac_time}) reduces to
\begin{eqnarray}\label{eq:ZVS}
C = 2 U - 2 (1 + sw) \, \dfrac{\beta m_1}{c} \, \arctan \left( \dfrac{y - y_1}{x - x_1} \right). 
\end{eqnarray} 

Therefore, Eq. \eqref{eq:ZVS} defines the zero velocity surfaces of the problem, or in other words, it determines the forbidden regions of motion for a test particle. Note that for $\beta=0$, the expression for ZVS coincides with the usual expression for the classical ERFBP.

Finally, concerning the stability of the fixed points, it is a well-known fact that detailed information about the motion of a system can be obtained by linearizing the equations of motion about a fixed point, $(x^{*}, y^{*})$. For this, we first introduce the following changes of variable: 
$\dot{x} \rightarrow \xi$, and $\dot{y}\rightarrow \eta$, such that the equations of motion (\ref{eqm1}-\ref{eqm2}) can be written as the first-order system
\begin{eqnarray}
\dot{x} &=& \xi \label{eq:ls1}\\
\dot{y} &=& \eta \label{eq:ls2}\\
\dot{\xi} &=& \frac{\partial U(x,y)}{\partial x} +2\eta- (1 + sw) F_{x}(x,y,\xi,\eta) \label{eq:ls3}\\
\dot{\eta} &=& \frac{\partial U(x,y)}{\partial y}-2\xi - (1 + sw) F_{y}(x,y,\xi,\eta)
\label{eq:ls4}
\end{eqnarray}
Then, the system (\ref{eq:ls1}-\ref{eq:ls4}) can be linearized by means of the coefficient matrix
\begin{equation}
\mathbb{A} =
\left(
\begin{array}{cccc}
0 & 0 & 1 & 0\\
0 & 0 & 0 & 1\\
A_{11} & A_{12} & A_{13} & A_{14}\\
A_{21} & A_{22} & A_{23} & A_{24}
\end{array}
\right)
\end{equation}
with
\begin{eqnarray*}
A_{11} &=& \frac{\partial^2 U(x,y)}{\partial x^2}-(1 + sw) \frac{\partial F_{x}(x,y,\xi,\eta)}{\partial x} ,\\
A_{12} &=&\frac{\partial^2 U(x,y)}{\partial x\partial y}-(1 + sw) \frac{\partial F_{x}(x,y,\xi,\eta)}{\partial y} ,\\
A_{13} &=& -(1 + sw) \frac{\partial F_{x}(x,y,\xi,\eta)}{\partial \xi} ,\\
A_{14} &=& 2-(1 + sw) \frac{\partial F_{x}(x,y,\xi,\eta)}{\partial \eta} ,\\
A_{21} &=& \frac{\partial^2 U(x,y)}{\partial x\partial y}-(1 + sw) \frac{\partial F_{y}(x,y,\xi,\eta)}{\partial x} ,\\
A_{22} &=& \frac{\partial^2 U(x,y)}{\partial y^2}-(1 + sw) \frac{\partial F_{y}(x,y,\xi,\eta)}{\partial y} , \\
A_{23} &=&-2-(1 + sw) \frac{\partial F_{y}(x,y,\xi,\eta)}{\partial \xi} ,\\
A_{24} &=&-(1 + sw) \frac{\partial F_{y}(x,y,\xi,\eta)}{\partial \eta} ,
\end{eqnarray*}
and therefore, the characteristic polynomial will be given by the quartic equation
\begin{eqnarray}
\nonumber &\lambda ^4& - \lambda ^3(A_{13} + A_{24})\\
\nonumber &+& \lambda ^2(A_{13}\,A_{24} - A_{14}\,A_{23} - A_{22} - A_{11})\\
\nonumber &+& \lambda (A_{11}\,A_{24} + A_{13}\,A_{22} - A_{12}\,A_{23} - A_{14}\,A_{21})\\
&+& A_{11}\,A_{22} - A_{12}\,A_{21} = 0.
\label{eq:charac_eq}
\end{eqnarray}
The equilibrium points are stable if all the roots of the characteristic polynomial evaluated at $(x^{*},y^{*},0,0)$ are pure imaginary roots or complex roots with negative real parts; otherwise, they are unstable.

%%%%%%%%%%%%%%%%%%%%%%%%%%%%%%%%%%%%%%%%%%%%%%
\subsection{Case 1: $m_1=m_2=m_3$}
\label{ssec1}
%%%%%%%%%%%%%%%%%%%%%%%%%%%%%%%%%%%%%%%%%%%%%%
Let us start considering the case in which the three primary bodies have the same value of mass, {\it i.e}, $m_1=m_2=m_3=1/3$. According to Eqs. \eqref{eq:coordinates}, the coordinates of the primaries $(x_1, y_1), (x_2, y_2),$ and $(x_3, y_3)$, are given respectively by 
\begin{equation*}
\left(\frac{1}{\sqrt{3}},0\right),   \left(-\frac{1}{2\sqrt{3}}, \frac{1}{2}\right), \quad {\rm and} \quad  \left(-\frac{1}{2\sqrt{3}}, -\frac{1}{2}\right). 
\end{equation*}

\begin{figure}
\centering
\includegraphics[width = \columnwidth]{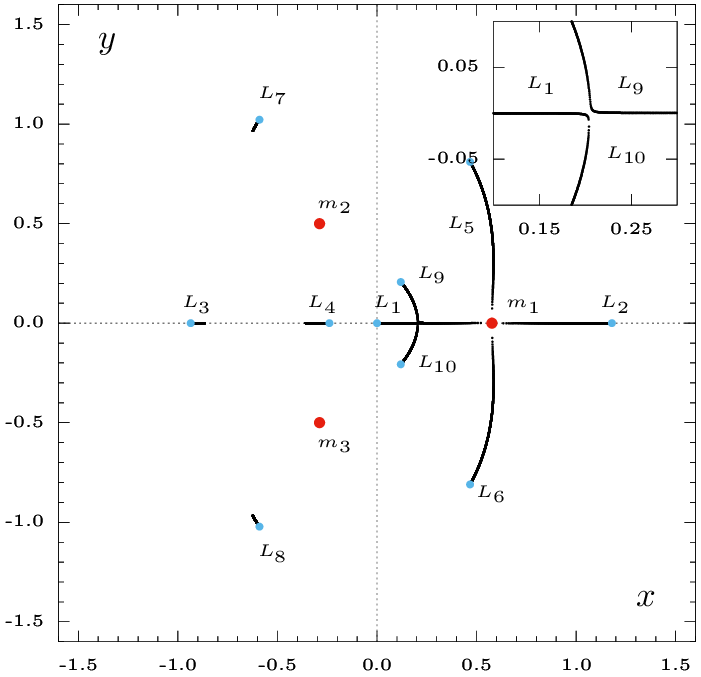}
\caption{(Color online). Parametric evolution of the libration points for $\beta \in \left[0, 1\right]$. Blue dots indicate the location of the libration points for $\beta=0$, while red dots represent the coordinates of the primaries.} 
\label{fig1}
\end{figure}

In Fig. \ref{fig1}, we have plotted the location and evolution of the equilibrium points for $\beta$ increasing from $0$ to  $1$. Blue dots are the solutions for $\beta = 0$, while red dots denote the coordinates of the three massive bodies\footnote{Along the paper, we shall use the customary value for the ratio of solar wind to Poynting-Robertson drag, $sw=0.35$.}. It can be seen that as $\beta$ increases, $L_1$ and $L_{10}$ reach both the same point along the $x$-axis and completely disappear for $\beta\approx 0.687$.  Also, it is observed that the libration points $L_2$, $L_5$, $L_6$,  and $L_9$, move gradually toward the position of the radiating body until they disappear for $\beta\approx 0.999$. On the other hand, $L_3$, $L_4$, $L_7$, and $L_8$, exhibit a small displacement from their initial positions (see Table \ref{table2} for a detailed description).

\begin{figure*}[t!]
\centering
\includegraphics[width = \linewidth]{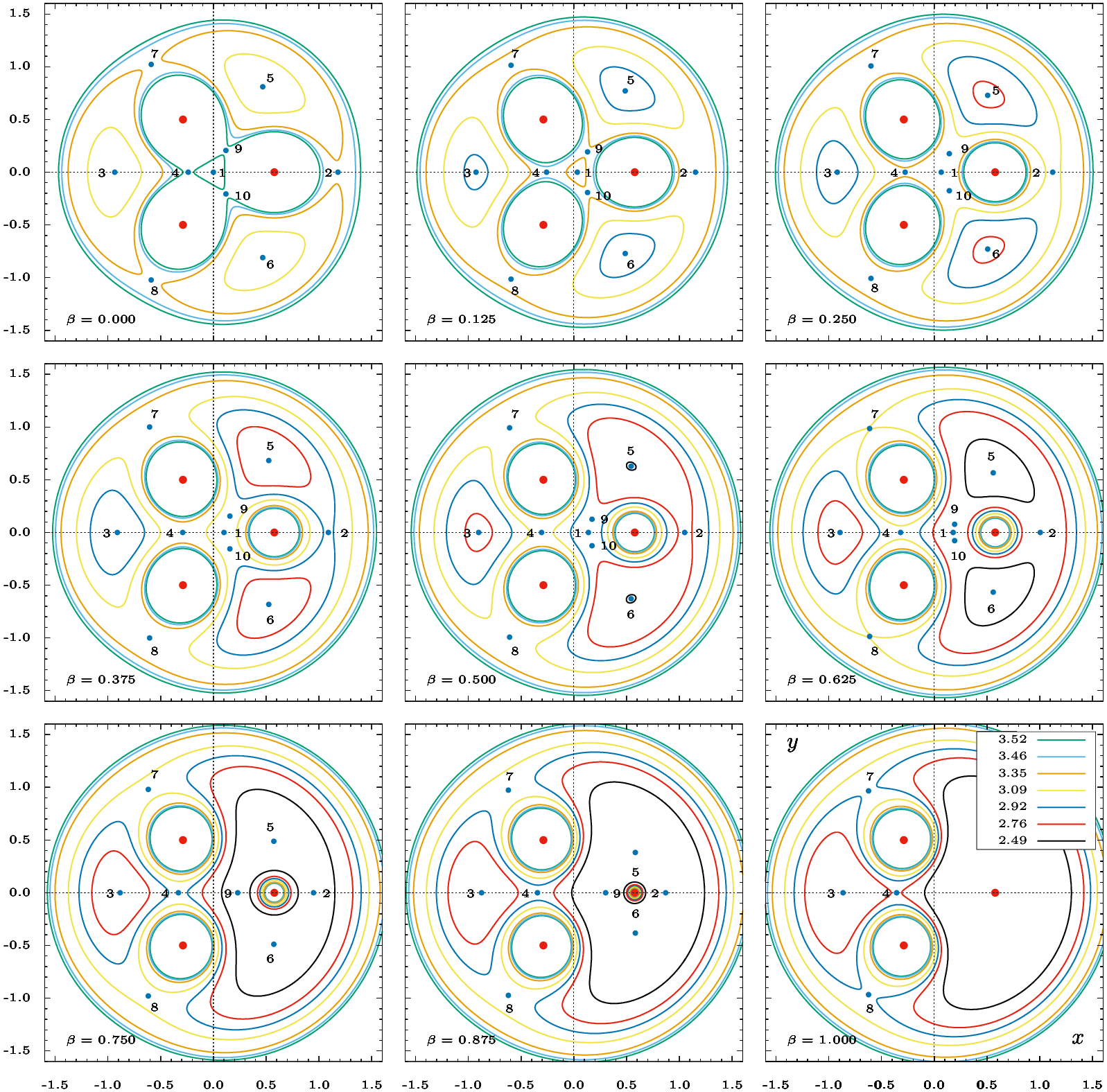}
\caption{(Color online). Zero velocity surfaces for increasing values of the radiation factor $\beta$, using $m_1 = m_2 = m_3 = 1/3$. Blue dots indicate the location of the libration points, while red dots denote the coordinates of the primaries. The color code is indicated in the lower-right panel.} \label{fig2}
\end{figure*}

\begin{figure*}[t!]
\centering
\includegraphics[width = \linewidth]{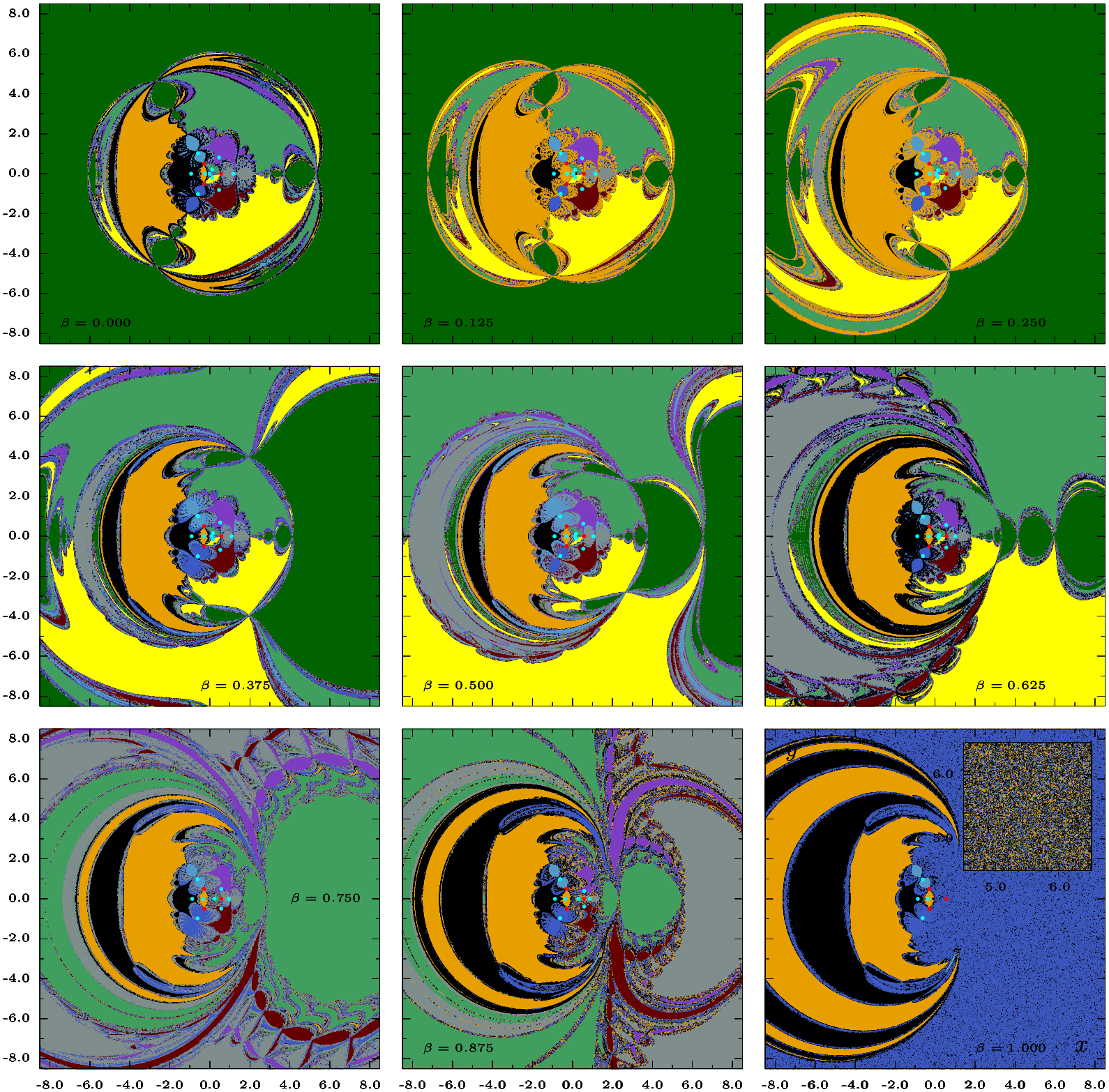}
\caption{(Color online). Basins of convergence using $m_1 = m_2 = m_3 = 1/3$, for increasing values of the radiation factor $\beta$. Cyan dots indicate the location of the libration points, while red dots denote the coordinates of the primaries. The color code is specified in the text.}
\label{fig3}
\end{figure*}

On the subject of the stability of the fixed points, it is found that in accordance with Ref. \cite{Baltagiannis2011}, for $\beta = 0$, all the equilibrium points are unstable. In the interval $\beta \in (0,1]$, the stability of the fixed points remains unaltered, since the form of the roots for all the equilibria does not change, {\it i.e.}, for $L_1$, $L_3$, $L_5$ and $L_6$, the characteristic equation (\ref{eq:charac_eq}) gives place to complex eigenvalues of the form $\lambda_{1,2,3,4} = \pm a \pm ib$, while for the libration points $L_2$, $L_4$, $L_7$, $L_8$, $L_9$ and $L_{10}$, all roots takes the form $\lambda_{1,2} = \pm \, ib$ and $\lambda_{3,4} = \pm \, a$. So, we may conclude that if the primary bodies are equal mass, the stability of the libration points does not change with the radiation factor $\beta$.  

\begin{table}[t!]
\centering
{\begin{tabular}{|ccc|}
\hline
{\text{Interval}} & {N. Equilibria} & {\text{Equilibria}} \\ %\cellcolor[rgb]{0.33,0.41,0.58}\textcolor{white}
\hline\hline
$\beta \in [0.000, 0.687]$ & 10 & $L_{1,2,3,4,5,6,7,8,9,10}$ \\
$\beta \in [0.688, 0.999]$ &  8 & $L_{2,3,4,5,6,7,8,9}$\\
$\beta = 1.000$ & 4 & $L_{3,4,7,8}$ 
\\\hline
\end{tabular}}
\caption{Existence of equilibrium points with the variation of the radiation factor $\beta$, for $m_1= m_2 = m_3$. \label{table2}}
\end{table}

Moreover, in Fig. \ref{fig2}, we show the parametric evolution of the ZVS with $\beta$, for different values of the Jacobi constant. Here, it can be noted that the regions of allowed motion are substantially modified for larger values of $\beta$. For example, when $C = 2.92$ (blue contour) and for $\beta = 0.125$, the forbidden region is made up of three small islands located just around the libration points $L_3$, $L_5$, and $L_6$. Then, the test particle will be free to move almost without any restriction around the primary bodies. However, as $\beta$ increases, these small islands get bigger, firstly, forming a small connection near $m_1$ until completely surround it. For $\beta\rightarrow 1$, the forbidden region in left-hand-side joins the one surrounding $m_1$, until the allowed regions of motion around $m_2$ and $m_3$, become almost isolated. This effect can be explained if we consider that, $\beta\approx 1$, implies that the force due to radiation pressure is comparable to the force of gravity, being able to push the test particle away from $m_1$. 

We finish this subsection by discussing the so-called basins of convergence and their evolution with the radiation factor $\beta$. As pointed out in section \ref{intro}, by basin of convergence, we refer to the set of points that after successive iterations converge to a specific fixed point. The numerical procedure for obtaining such basins is the multivariate Newton-Raphson method, which can be defined by the map
\begin{equation}
{\bf{x}}_{n+1} = {\bf{x}}_{n} - J^{-1}f({\bf{x}}_{n}),
\label{sch}
\end{equation}
where ${\bf{x}}=(x,y)$, $f({\bf{x_n}})$ represents the system of equations (\ref{eq:equil_x}-\ref{eq:equil_y}), and $J^{-1}$ is the inverse Jacobian matrix. The initial guesses are defined within the region enclosed  by $x \in \left[-8,5; \, 8,5\right]$ and $y \in \left[ -8,5; \, 8,5\right]$, the step size is equal to $10^{-2}$, and tolerance of the order $10^{-12}$. Each initial condition leading to a given fixed point is plotted using the following color code: $L_1 \rightarrow$ dark green,  $ L_2 \rightarrow$ dark grey,  $L_3 \rightarrow$ black,  $L_4 \rightarrow$ orange,  $L_5 \rightarrow$ purple, $L_6 \rightarrow$ red wine, $L_7 \rightarrow$ light blue, $L_8 \rightarrow$ blue,  $L_9 \rightarrow$ green and  $L_{10} \rightarrow$ yellow; while the libration points are denoted by cyan dots.

In Fig. \ref{fig3}, we plot the basins of convergence of the equilibrium points, for nine different values of the radiation factor $\beta$. Here, it can be easily observed that some of the equilibrium points disappear as $\beta$ increases, or in other words, the number of colors for $\beta=1$ is reduced compared to the case $\beta=0$. It should be noted that as soon as $\beta >0$ the $2\pi/3$ symmetry is broken and replaced by a reflection symmetry with respect to the $x$-axis.  Additionally, the complexity of the basins increases for higher values of the radiation parameter, despite the fact that the number of libration points is much smaller than in the case $\beta=0$. The last statement can be clearly envisioned in the last panel of Fig. \ref{fig3}, where the right-hand side region becomes very noisy, suggesting an increasing unpredictability of the basins.

%%%%%%%%%%%%%%%%%%%%%%%%%%%%%%%%%%%%%%%%%%%%%%
\subsection{Case 2: $m_1\neq m_2=m_3$}
\label{ssec2}
%%%%%%%%%%%%%%%%%%%%%%%%%%%%%%%%%%%%%%%%%%%%%%
As a second case, we have assumed that $m_2 = m_3={\mathfrak{m}}$. Accordingly, the mass parameter of the radiating body will be given by $m_1 = 1 - 2  {\mathfrak{m}}$, and the coordinates of the primaries $(x_1, y_1), (x_2, y_2),$ and $(x_3, y_3)$, read as
\begin{equation*}
\left(\sqrt{3} {\mathfrak{m}}, 0\right),   \left(\frac{\sqrt{3}}{2}(2 {\mathfrak{m}}-1), \frac{1}{2}\right), \,\, {\rm and} \,\,  \left(\frac{\sqrt{3}}{2}(2 {\mathfrak{m}}-1), -\frac{1}{2}\right). 
\end{equation*}

\begin{table}[t!]
\centering
{\begin{tabular}{|ccc|}
\hline
{${\mathfrak{m}}$} & {\text{Interval}} & {\text{Equilibria}} \\ %\cellcolor[rgb]{0.33,0.41,0.58}\textcolor{white}
\hline\hline
\multirow{3}{*}{$0.05$}& $\beta \in [0.000, 0.928]$ & $L_{1,2,5,6,7,8,9,10}$ \\
& $\beta \in [0.929, 0.999]$ & $L_{2,5,6,7,8,9}$\\
& $\beta = 1.000$ & $L_{7,8}$ \\\hline
\multirow{3}{*}{$0.10$} & $\beta \in [0.000, 0.912]$ & $L_{1,2,5,6,7,8,9,10}$ \\
& $\beta \in [0.913, 0.999]$ & $L_{2,5,6,7,8,9}$ \\ 
& $\beta = 1.000$ & $L_{7,8}$ \\\hline
\multirow{5}{*}{$0.15$} & $\beta \in [0.000, 0.350]$ & $L_{2,3,5,6,7,8,9,10}$ \\
& $\beta \in [0.351, 0.360]$  & $L_{1,2,3,4,5,6,7,8,9,10}$ \\ 
& $\beta \in [0.361, 0.889]$  & $L_{1,2,5,6,7,8,9,10}$ \\
& $\beta \in [0.890, 0.999]$  & $L_{2,5,6,7,8,9}$ \\
& $\beta = 1.000$ & $L_{7,8}$ \\\hline
\multirow{5}{*}{$0.20$} & $\beta \in [0.000, 0.293]$ & $L_{2,3,5,6,7,8,9,10}$ \\
& $\beta \in [0.294, 0.608]$  & $L_{1,2,3,4,5,6,7,8,9,10}$ \\ 
& $\beta \in [0.609, 0.859]$ & $L_{1,2,5,6,7,8,9,10}$ \\
& $\beta \in [0.860, 0.999]$  & $L_{2,5,6,7,8,9}$ \\
& $\beta = 1.000$  & $L_{7,8}$ \\\hline
\multirow{4}{*}{$0.25$} & $\beta \in [0.000, 0.160]$ & $L_{2,3,5,6,7,8,9,10}$ \\
& $\beta \in [0.161, 0.816]$  & $L_{1,2,3,4,5,6,7,8,9,10}$ \\ 
& $\beta \in [0.817, 0.999]$  & $L_{2,3,4,5,6,7,8,9}$ \\
& $\beta = 1.000$ & $L_{3,4,7,8}$ \\\hline
\multirow{3}{*}{$0.30$} & $\beta \in [0.000, 0.751]$ & $L_{1,2,3,4,5,6,7,8,9,10}$ \\
& $\beta \in [0.752, 0.999]$ & $L_{2,3,4,5,6,7,8,9}$ \\ 
& $\beta = 1.000$ & $L_{3,4,7,8}$ \\\hline
\multirow{3}{*}{$0.35$} & $\beta \in [0.000, 0.644]$ & $L_{1,2,3,4,5,6,7,8,9,10}$ \\
& $\beta \in [0.645, 0.999]$ & $L_{2,3,4,5,6,7,8,9}$ \\ 
& $\beta = 1.000$ & $L_{3,4,7,8}$ \\\hline
\multirow{3}{*}{$0.40$} & $\beta \in [0.000, 0.430]$ & $L_{1,2,3,4,5,6,7,8,9,10}$ \\
& $\beta \in [0.431, 0.999]$ & $L_{2,3,4,5,6,7,8,9}$ \\ 
& $\beta = 1.000$ & $L_{3,4,7,8}$ \\\hline
\multirow{2}{*}{$0.45$} & $\beta \in [0.000, 0.999]$ & $L_{1,2,3,4,5,6,7,8}$ \\
& $\beta = 1.000$ & $L_{3,4,7,8}$ \\\hline
\end{tabular}}
\caption{Existence of equilibrium points with the variation of the radiation factor $\beta$, for different values of $m_2 = m_3$. \label{table3}}
\end{table}

The last result implies that not only the position of the primaries depends on the parameter  ${\mathfrak{m}}$, but the existence and location of libration points will also depend on this parameter and on $\beta$, according to Eqs. (\ref{eq:equil_x}, \ref{eq:equil_y}). In Table \ref{table3}, we present how the total number of equilibrium points varies on different intervals of the radiation factor and for different values of ${\mathfrak{m}}$. It can be observed that for larger values of $\beta$ the total number of fixed points is reduced, where the final number of surviving points increases for large values of ${\mathfrak{m}}$.

\begin{figure*}[t!]
\centering
\includegraphics[width = \linewidth]{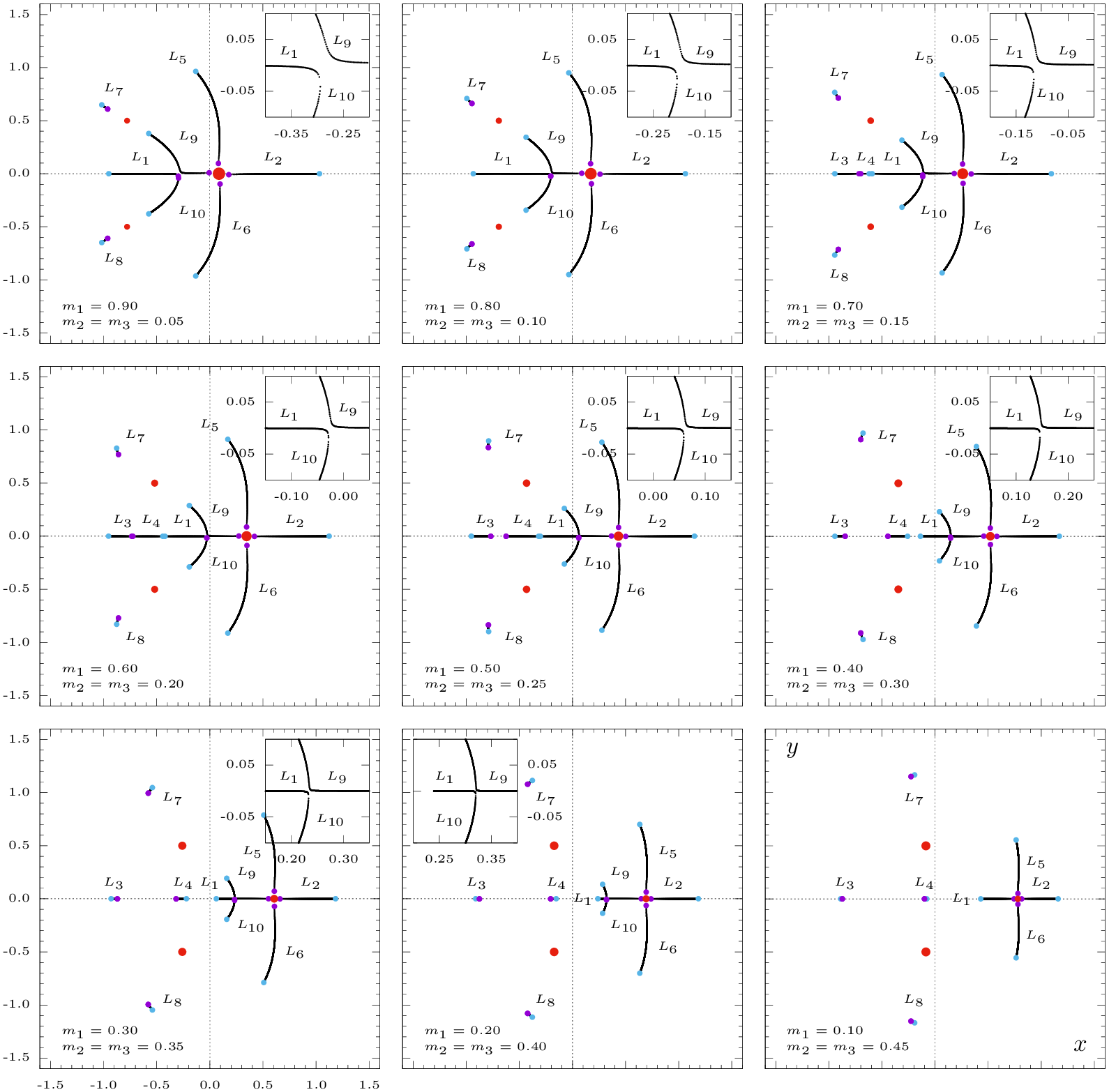}
\caption{(Color online). Existence and evolution of the equilibrium points with the variation of the radiation factor $\beta$, for different cases of $m_2 = m_3$. Cyan dots denote the initial position of the equilibria, while purple dots show their final position. The location of the primaries is indicated in red color.}%% no full stop at the end of caption
\label{fig4}
\end{figure*}

\begin{figure*}[t!]
\centering
\includegraphics[width = \linewidth]{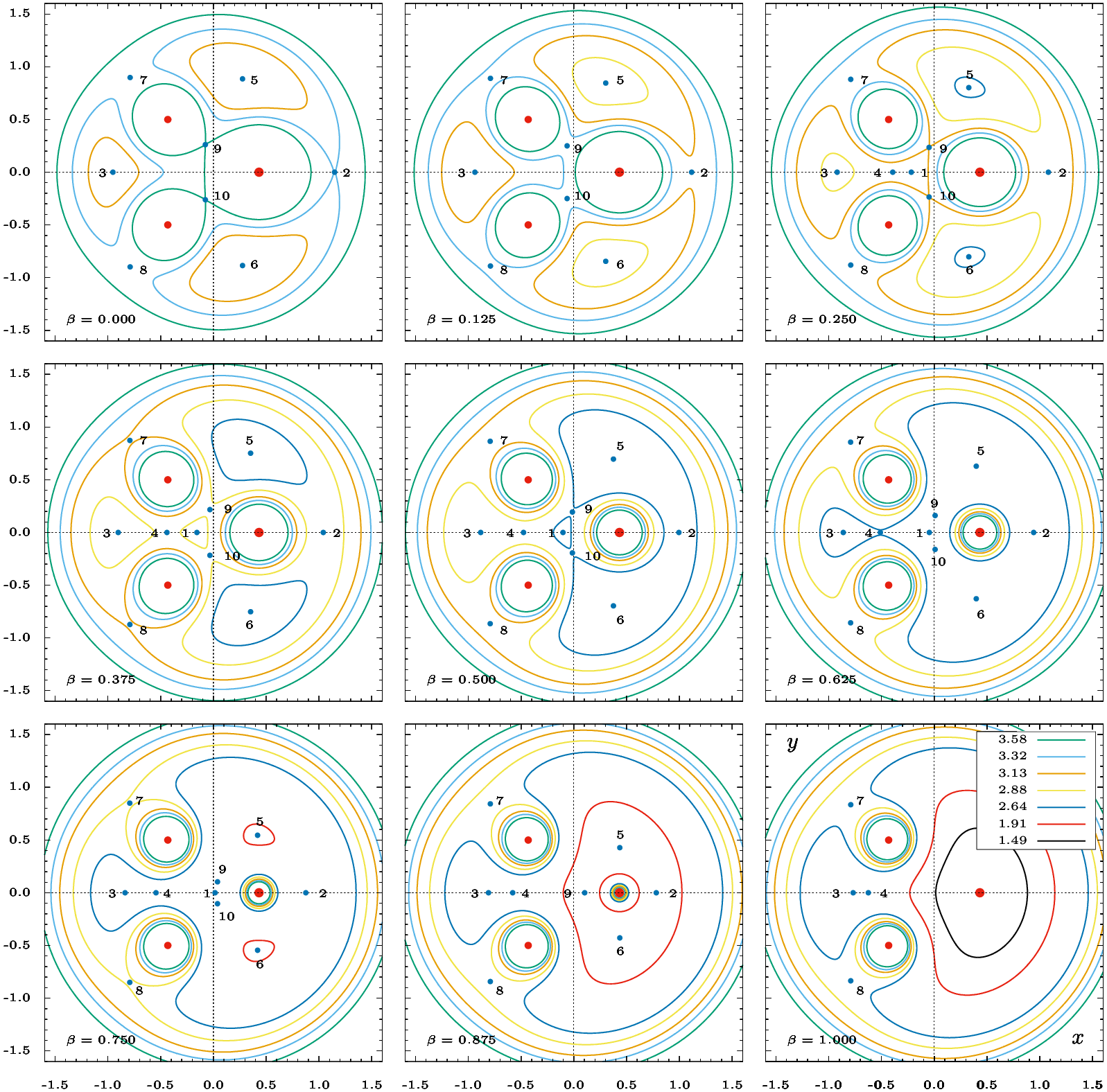}
\caption{(Color online). Zero velocity surfaces for different values of the radiation factor $\beta$, when $m_1 = 0.5$ and $m_2 = m_3 = 0.25$. Blue dots indicate the location of the libration points, while red dots denote the position of the primaries. The color code is indicated in the lower-right panel.} %% no full stop at the end of caption
\label{fig5}
\end{figure*}

In Figure \ref{fig4}, we present a graphic representation of the evolution of equilibrium points with the variation of the radiation factor $\beta$, for different values of $m_2 = m_3={\mathfrak{m}}$. This figure can be analyzed as follows: {\it (i)} in the case $m_1\gg {\mathfrak{m}}$, it can be noted that the collinear points $L_3$ and $L_4$ do not exist. For larger values of $\beta$, the equilibria $L_1$ and $L_{10}$  approximate each other until they join near the $x$-axis, while the points $L_2$, $L_5$, $L_6$, and $L_9$ move toward the radiating body until they disappear for $\beta=1$. {\it (ii)} As expected, when $m_1\approx {\mathfrak{m}}$ the results are exactly like the ones described in the previous subsection. {\it (iii)} For $m_1\ll {\mathfrak{m}}\approx 0.5$, the non-collinear points $L_9$ and $L_{10}$ do not exist. As $\beta$ increases, $L_3$, $L_4$, $L_7$, and $L_8$ exhibit very small displacements with respect to their initial positions, while the libration points $L_1$, $L_2$, $L_5$, and $L_6$ move abruptly toward the radiating body, until they disappear for $\beta=1$.

Concerning the stability of the fixed points, we start considering the case $\beta = 0$. Here, we found that if $\mathfrak{m} \in (0, 0.0027]$ the libration points $L_3$, $L_5$, and $L_6$ are stable, for $\mathfrak{m} \in [0.0027, 0.0188]$,  only $L_5$, and $L_6$ remain stable, but for  $\mathfrak{m}>0.0188$ all the equilibria are unstable. All those results agree with the ones given in Ref. \cite{Baltagiannis2011}. Once we start varying $\beta$ in steps of $\Delta \beta = 1 \times 10^{-3}$, we get: (i) for $\beta = 0.001$ and $\mathfrak{m} \in (0,0.0024]$ the libration points $L_3$, $L_5$ and $L_6$ are stable; for $\mathfrak{m} \in [0.0025,0.0185]$ only $L_5$ and $L_6$ are stable, while for $\mathfrak{m} > 0.0185$ all the equilibria are unstable. (ii) For $\beta = 0.002$ and $\mathfrak{m} \in (0,0.0011]$ the fixed points $L_3$, $L_5$ and $L_6$ are stable; for $\mathfrak{m} \in [0.0012,0.0174]$ only $L_5$ and $L_6$ are stable, but for $\mathfrak{m} > 0.0174$ all the fixed points are unstable. (iii) For $\beta = 0.003$ and $\mathfrak{m} \in (0,0.0147]$ only $L_5$ and $L_6$ are stable, while for $\mathfrak{m} > 0.0147$ all the equilibria are unstable. (iv) For $\beta=0.004$ and $\mathfrak{m} \in (0,0.0093]$ $L_5$ and $L_6$ are stable, but for $\mathfrak{m} > 0.0093$ all the fixed points are unstable. (v) Finally, for $\beta > 0.004$, all fixed points are unstable regardless the value of $\mathfrak{m}$.

From the previous results, we can infer that the stability of the libration points remains unaltered, only for values of radiation factor of the order $10^{-3}$, however, for larger values of $\beta$, all the equilibria become unstable. At this point, it should be noted that the radiation factor of the solar system is of the order $10^{-1}$, which implies that, in a realistic system, such factor should be able to destroy the stability of the fixed points. 

On the other hand, in Figure \ref{fig5}, we present the ZVS by using different values of the Jacobi constant. The tendency in this figure is very similar to the one observed in Fig. \ref{fig2}. For example, in the case $C=2.64$ and for small values of $\beta$, say $\beta < 0.25$, the test particle is free to move in the whole configuration space, however, as $\beta$ increases the libration points $L_5$ and $L_6$ become isolated points. Such islands start growing in size until  the ZVS surrounds completely $m_1$, in particular, for $\beta=0.875$ the motion of the test particle is limited to a small region around $m_1$, or to a larger zone that allows collisions with $m_2$ and $m_3$, but imposes a barrier in the vicinity of $m_1$. This behavior is consistent with the idea that, for $\beta\rightarrow 1$, the drag forces are strong enough to push away the test particle from $m_1$.

\begin{figure*}[t!]
\centering
\includegraphics[width = \linewidth]{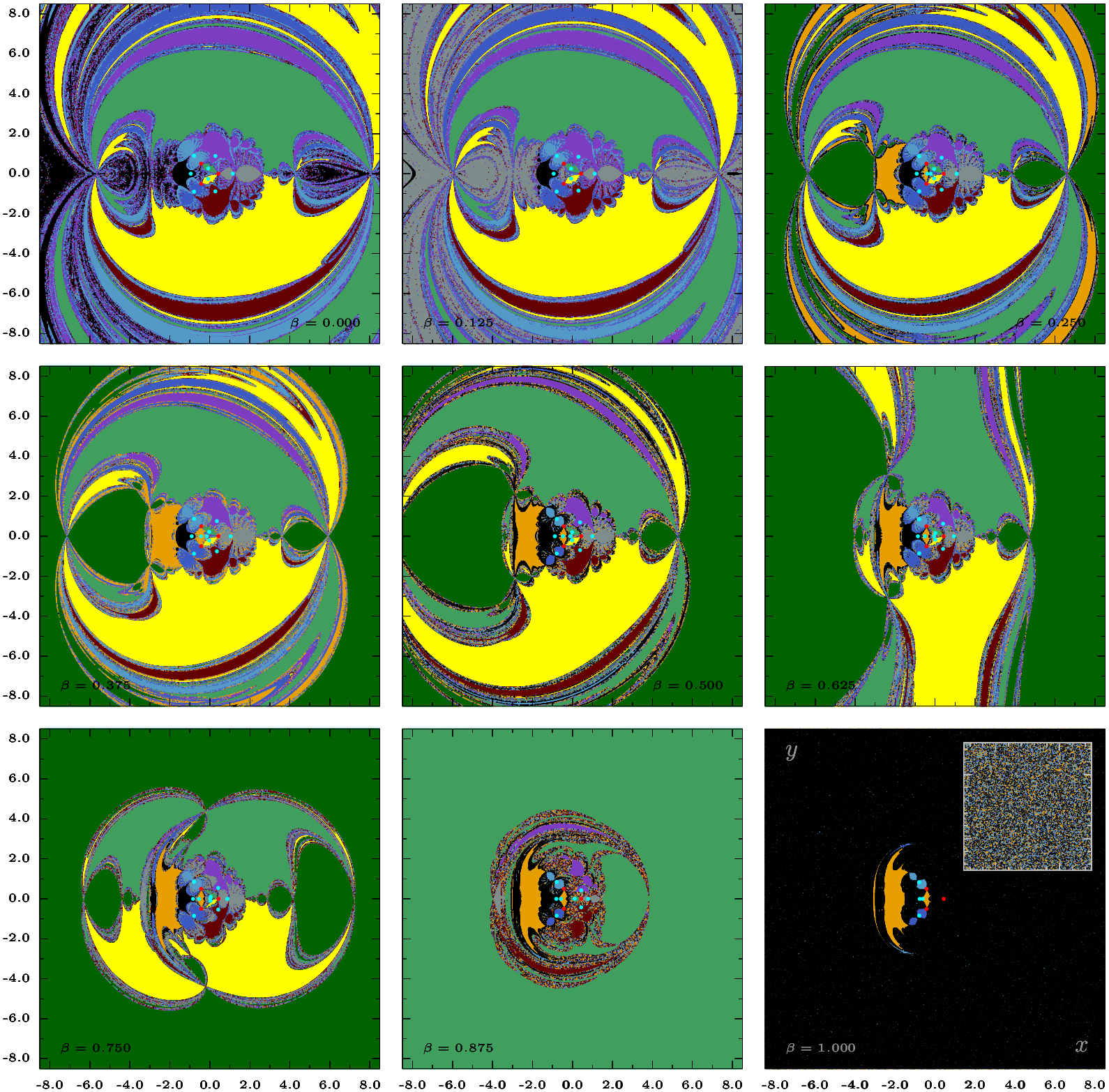}
\caption{(Color online). Basins of attraction for $m_1 = 0.5$, $m_2 = m_3 = 0.25$, for increasing values of the radiation factor $\beta$. Cyan dots indicate the location of the libration points, while red dots denote the coordinates of the primaries. The color code is the same used figure \ref{fig3}.}
\label{fig6}
\end{figure*}

In Figure \ref{fig6} we use the same color code of Fig. \ref{fig3}. Here, it can be noted that in the case $m_1 = 0.5$, $m_2 = m_3 = 0.25$, the basins of attraction exhibit a very complex pattern symmetric in shape, but asymmetric in color. Due to the fact that the number of fixed points grows for intermediate values of $\beta$ ($\beta \in [0.25, 0.75]$) and decreases for larger values of the radiation factor ($\beta>0.875$), the highly fractal structures are always present. Related to the extent of the basins of convergence, it is seen that for $\beta\in[0, 0.375]$ the area corresponding to libration point $L_9$ and $L_{10}$ predominates, while for  $\beta\in[0.5, 0.75]$ the area of the set of points converging to the fixed point $L_1$ is bigger than the one for $L_9$ and $L_{10}$. Special attention deserves the case $\beta=0.625$, where the basins seem stretched, here there are three prevailing colors, the ones corresponding to $L_1, L_9$ and $L_{10}$, i.e., the fixed points that are closer to the origin of coordinates. Such a structure allows us to prognosticate a reduction of the fractality of the basin. 

Finally, for $\beta>0.875$, the area corresponding to the basin for $L_9$ increases, while for $\beta=1$ it seems to be dominated by $L_3$, however, in the inset of the lower right panel, it is observed that the black region is, in fact, a very noisy region composed of initial conditions tending to all the libration points. From the previous description, we may infer that in the case of two equal masses, the determinations of the final state via the root finding algorithm of Newton-Raphson is highly sensitive, mainly, for lower and higher values of $\beta$.

\subsection{Case 3: $m_1\neq m_2 \neq m_3$}
\label{ssec3}

As a last scenario, we consider the equilateral triangle configuration Sun-Jupiter-Trojan Asteroid, in addition with the P-R and solar wind drag forces. In the normalized units, the masses of the primaries are given by  $m_1 = m_{S} = 0.999046321943$, $m_2 = m_{J} = 0.000953678050$ and $m_3 = m_{A} = 6.99996 \times 10^{-12}$, where $m_3$ corresponds to the mass of 624 Hektor, an actual asteroid of the Trojan group\footnote{Unlike the previous cases, all calculations of this subsection were performed in quadruple precision and the speed of light was set as in Baltagiannis \cite{Baltagiannis2013}}. 

According with Eqs. (\ref{eq:coordinates}), the coordinates of the primary bodies are now
\begin{eqnarray*}
(x_1, y_1)&\approx &(0.000953678, 0.000000),\\  
(x_2, y_2)&\approx &(-0.999046, 6.35659\times10^{-9}), \\ 
(x_3, y_3)&\approx &(-0.499046, -0.866025).
\end{eqnarray*}

\begin{figure}
\centering
\includegraphics[width = \columnwidth]{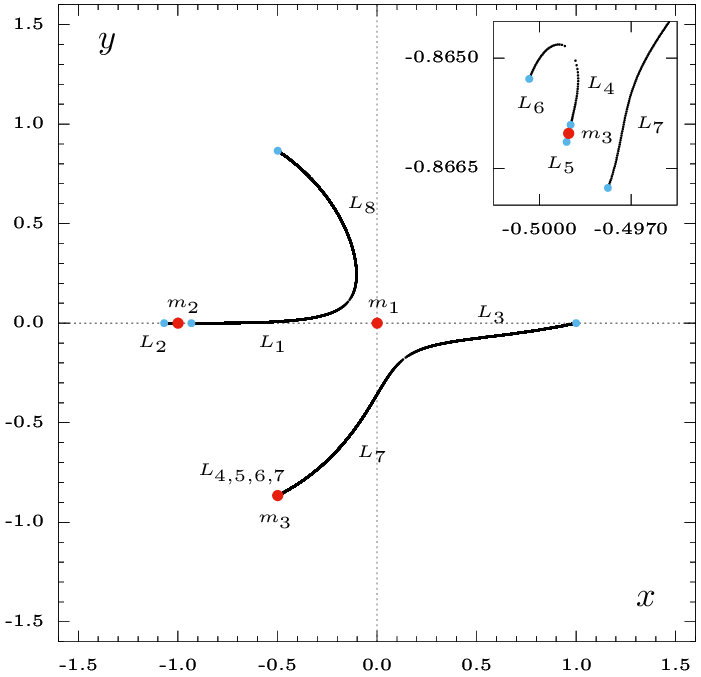}
\caption{(Color online). Parametric evolution of the libration points for $\beta \in \left[0, 1\right]$. Cyan dots indicate the location of the libration points for $\beta=0$, while red dots represent the coordinates of the primaries. In the upper right corner, we show a zoom of the parametric evolution around $m_3$.} 
\label{fig7}
\end{figure}

In agreement with the seminal paper by  Baltagiannis \cite{Baltagiannis2013}, in absence of radiation pressure and drag forces ($\beta = 0$), the Sun-Jupiter-Trojan Asteroid system admits eight non-collinear equilibrium points, three of them ($L_6$, $L_7$, and $L_8$) linearly stable. The adjective non-collinear is due to the fact that $L_1$, $L_2$, and $L_3$,  do not lie exactly on the $x$-axis. As $\beta$ increases, the equilibria $L_4$ and $L_6$, reach both the same point and completely disappear for $0.0029 < \beta < 0.0030$. The same applies for $L_3$ and $L_7$ for $0.9887 < \beta < 0.9888$, and for $L_1 $ and $L_8$ for $0.9940 < \beta < 0.9941$ (we refer the reader to Table \ref{table4} for details). On the other hand, $L_2$ and $L_5$ hardly move, approaching each one to its nearest primary (See Fig. \ref{fig7}).

\begin{table}[t!]
\centering
{\begin{tabular}{|ccc|}
\hline
{\text{Interval}} & {N. Equilibria} & {\text{Equilibria}} \\ %\cellcolor[rgb]{0.33,0.41,0.58}\textcolor{white}
\hline\hline
$\beta \in [0.0000, 0.0029]$ & 8 & $L_{1,2,3,4,5,6,7,8}$ \\
$\beta \in [0.0030, 0.9887]$ &  6 & $L_{1,2,3,5,7,8}$\\
$\beta \in [0.9888, 0.9940]$ & 4 & $L_{1,2,5,8,}$ \\
$\beta \in [0.9941, 1.0000]$ &  2 & $L_{2,5}$\\
\hline
\end{tabular}}
\caption{Existence of equilibrium points with the variation of the radiation factor $\beta$, for the Sun-Jupiter-Trojan Asteroid system. \label{table4}}
\end{table}

\begin{figure*}[t!]
\centering
\includegraphics[width = \linewidth]{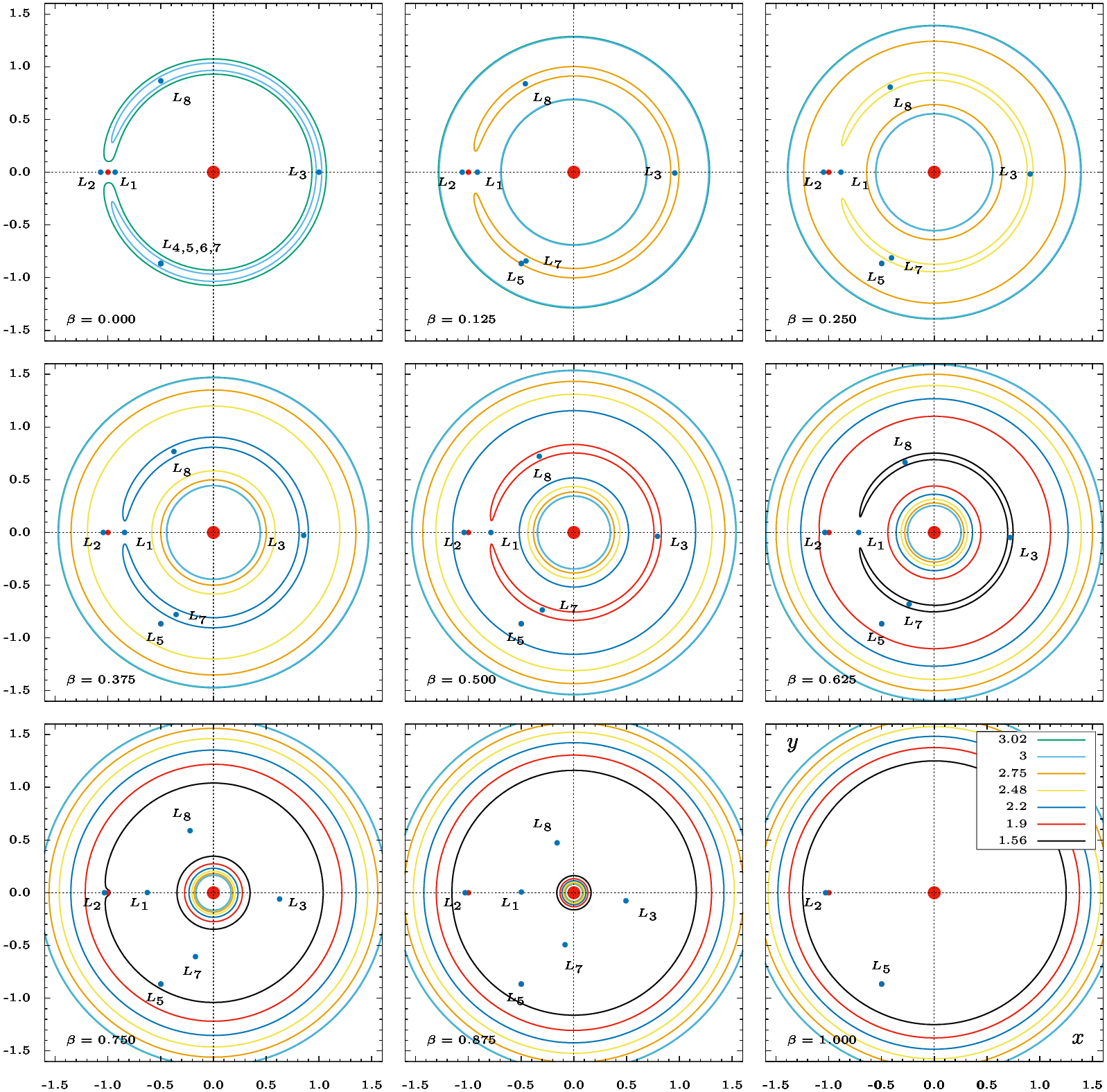}
\caption{(Color online). Zero velocity surfaces for increasing values of the radiation factor $\beta$, in the Sun-Jupiter-Trojan Asteroid-Spacecraft system. Blue dots indicate the location of the libration points, while red dots denote the position of the primaries. The color code is indicated in the lower-right panel.}
%% no full stop at the end of caption
\label{fig8}
\end{figure*}

On the stability subject, for $\beta \in (0,0.0029]$ the equilibria $L_6$, $L_7$, and $L_8$ are linearly stable, however, as $\beta$ grows the stability of the fixed points also changes, {\it i.e.},  for $\beta \in$ $[0.0030,0.0104]$, only $L_7$ and $L_8$ remain stable, while for $\beta  > 0.0104$, all the fixed points become unstable. Here, it can be noted that $\beta  \approx 10^{-2}$ corresponds to the largest value of the radiation factor (of the three considered cases) in which we still having stable fixed points. So, taking into account that $\beta\approx 10^{-1}$ in the Solar system, we may conclude that in the restricted four-body problem, Sun-Jupiter-Trojan Asteroid-Spacecraft (test particle), all the libration points must be unstable.

In Fig. \ref{fig8} we present the parametric evolution of the ZVS with $\beta$, for different values of the Jacobi constant. In absence of the P-R and drag forces, the forbidden regions of the Sun-Jupiter-Trojan Asteroid system are made of horseshoe-shaped contours, very similar to the ones observed in the Sun-Jupiter system in the circular restricted three-body problem (see {\it e.g.} \cite{Liou1995}). The circular horseshoe is open on $m_2$, such that this opening will be greater at lower values of $C$. Keeping a fixed value of $C$, as $\beta$ increases, the mentioned horseshoe closes on it selves forming a donut-shaped contour in which the allowed motion region becomes smaller and smaller around the radiating body. For larger values of $\beta$ ($\beta\rightarrow 1$), the circular region around $m_1$ disappears, and the forbidden region of motion becomes a circle containing the Lagrangian points $L_{2,5}$ and the three masses. Here, we would like to highlight that the location of $m_3$ almost coincides with the position of $L_5$, for this reason, the red dot of $m_3$ is not seen in Fig. \ref{fig8}.

\begin{figure*}[t!]
\centering
\includegraphics[width = \linewidth]{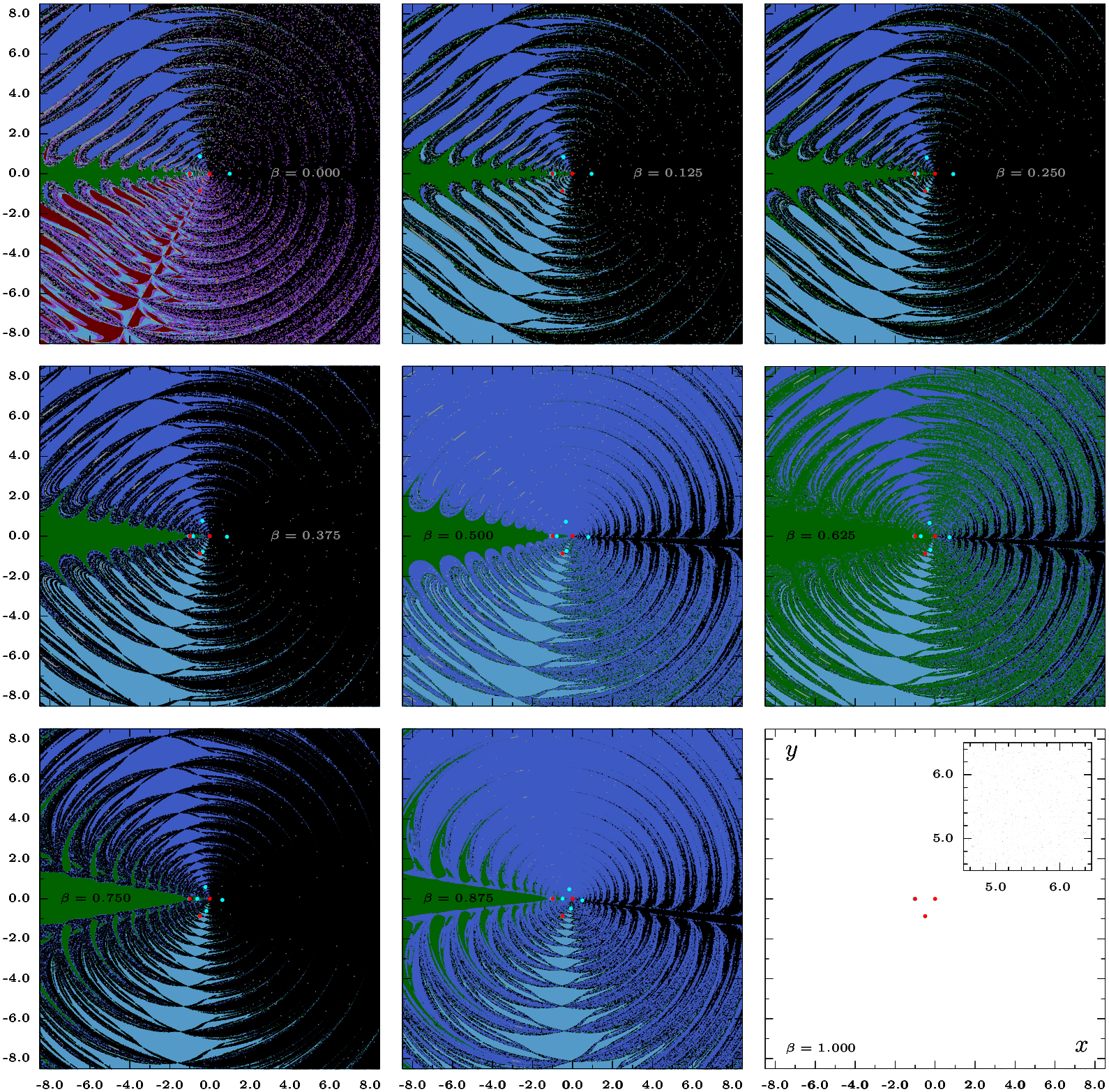}
\caption{(Color online). Basins of attraction in the Sun-Jupiter-Trojan Asteroid-Spacecraft system, for increasing values of the radiation factor $\beta$. Cyan dots indicate the location of the libration points, while red dots denote the coordinates of the primaries. The color code is the same used in figure \ref{fig3}.}%% no full stop at the end of caption
\label{fig9}
\end{figure*}

On the other hand, in Fig. \ref{fig9}, we plot the basins of convergence of the equilibrium points for the Sun-Jupiter-Trojan Asteroid system. The color code used in this case is the same as in figure \ref{fig3}. For $\beta = 0$, the frame exhibits the eight colors of the corresponding libration points, albeit the most noticeable basins are the ones corresponding to $L_1, L_3, L_5, L_6, L_7$, and $L_8$. In general, it is observed that along the negative $x-$axis the basin associated to $L_1$ predominates, while along the positive $x-$axis the one associated to $L_3$ does. In the remaining frames, the number of equilibria decreases to 6 ($L_1$, $L_2$, $L_3$, $L_5$, $L_7$, and $L_8$), then 4 ($L_1$, $L_2$, $L_5$, and $L_8$) and finally 2 ($L_2$ and $L_5$). However, due to the location of the equilibria $L_5$ (superimposed with $m_3$), no initial condition of our mesh converges to this point and therefore, the purple basin neither appears. 

In accordance with  cases \ref{ssec1} and \ref{ssec2}, the complexity of the basins increases for larger values of $\beta$, this is so because, for $\beta\rightarrow 1$,  $L_3$ (whose initial position starts alongside the x-axis) moves downward, while  $L_1$ moves upward, $L_7$ moves in counterclockwise and $L_8$ moves toward the radiating body $m_1$. Such displacements shall break the symmetry of the basins of convergence and hence increasing the intricacy of them (see {\it e.g.} the lower-mid panel in Fig. \ref{fig9}). Lastly, when $\beta\rightarrow 1$ we observe a practically white configuration space, this color is assigned to non-converging initial conditions, however, the inset of this panel shows a region composed by scattered dark gray dots. This results can be understood by considering that for $\beta=1$ there exist only 2 fixed points, one of them corresponds to $L_5$ which, due to its superposition with the primary $m_3$, do not allow for convergence of any initial condition of our mesh after $ 10^{5}$ iterations. 

\section{Basin entropy}
\label{sec4}

So far, the analysis of the basins of convergence has been purely qualitative, in order to make quantitative our study, in this section, we shall use a quantity recently introduced in \cite{Daza2016} which allow us to measure the uncertainty (complexity) of a given basin. The new dynamical quantity is termed basin entropy and provides a useful method to explore the differences observed in our system when the radiation and mass parameters are varied.

For the sake of completeness, we will briefly describe the idea behind the calculation of the basins' entropy. Assuming that the phase space contains $N_A$ different final states (or colors in our case), we divide the space of states into a grid of $N$ square cells, such that each one of these cells contains at least one of the $N_A$ states. Defining $p_{i,j}$ as the probability to detect a state $j$ in the $i-$th cell and by applying the Gibbs entropy definition to that set, the entropy for the $i-$th cell can be expressed as 
\begin{equation}
S_{i} = \sum_{j=1}^{N_{A}}p_{i,j}\log\left(\frac{1}{p_{i,j}}\right).
\label{si}
\end{equation}
Therefore, the basin entropy for the total number of cells $N$ in the basin is calculated as an average, {\it i.e.},
\begin{equation}
S_{b} = \frac{1}{N}\sum_{i=1}^{N} S_{i}=\frac{1}{N}\sum_{i=1}^{N} \sum_{j=1}^{N_{A}}p_{i,j}\log\left(\frac{1}{p_{i,j}}\right).
\label{sb}
\end{equation}

Strictly speaking, the average of this quantity must depend on the number of considered cells, such that for a larger value of $N$ the result for the basin entropy $S_b$ should be more precise. However, a bigger $N$ can be reached only with a smaller size of the square cells, which also have a minimum size to contain at least one of the $N_A$ states. To solve this issue, we follow the procedure outlined in \cite{Daza2017}, in which the square cells are randomly picked in the space of states through a Monte Carlo procedure, allowing us to increase the number of cells $N$ as necessary. In our particular problem, we find that the final value for the basin entropy keeps constant for a number of cells larger than $3\times10^{5}$, hence, in the three cases, we used $N=3.5 \times10^{5}$ cells.

\begin{figure}
\centering
\includegraphics[width = \columnwidth]{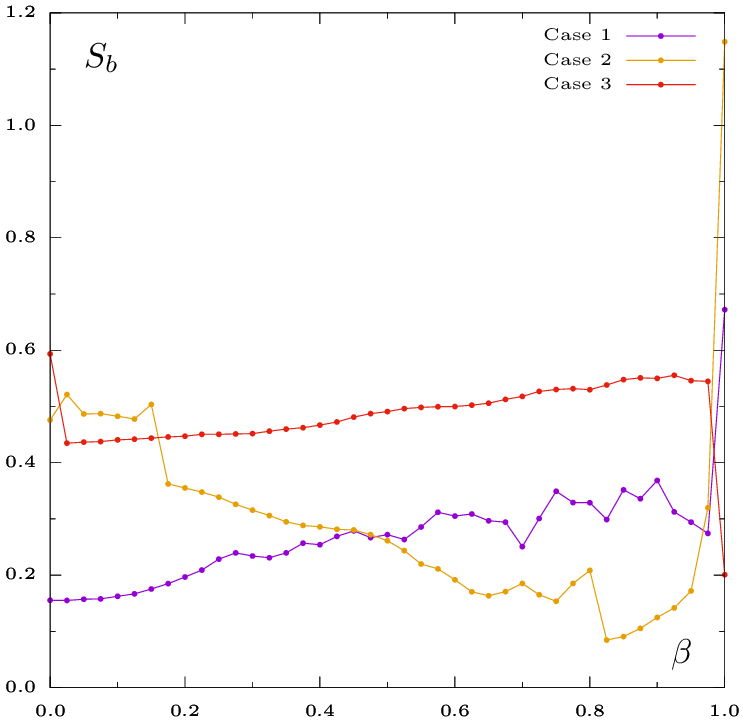}
\caption{(Color online). Basin entropy $S_{b}$ as a function of the radiation parameter $\beta$ in the three considered cases: (Case 1) $m_1 = m_2 = m_3$, (Case 2) $m_1 \ne m_2 = m_3$, and (Case 3) $m_1 \ne m_2 \ne m_3$. 
} 
\label{fig10}
\end{figure}

In Fig. \ref{fig10}, we present the parametric evolution of the basin entropy as a function of the radiation parameter $\beta$, for the considered cases. Our results indicate that, in case 1, the basin entropy increases almost monotonically with the radiation parameter $\beta$. A very similar behavior is observed for case 3 in the interval $\beta \in [0.025, 0.975]$,
yet for $\beta \in [0, 0.025]$ and $\beta \in [0.975, 1]$, the basin entropy decreases abruptly. Interestingly, the opposite tendency is observed for case 2, where the basin entropy decreases with the radiation parameter $\beta$ in the interval $\beta \in [0, 0.825]$, but increases in the interval $\beta \in [0.825, 1]$ from $S_b\approx 0.1$ up to a value of $S_b\approx 1.14$. In general, our findings suggest that in cases 1 and 2, the unpredictability associated to the NR basins of convergence for the triangular restricted four-body problem with an extreme radiating body ($\beta=1$) is larger in comparison with the non-radiating case. Nevertheless, in the case of Sun-Jupiter-Trojan Asteroid-Spacecraft system (case 3), the exact opposite situation occurs.  These results can be explained if we consider that in case 3, the final number of libration points is one-fourth of the initial points, {\it i.e.}, the number of final states $N_A$ is considerably diminished, and hence, the value of Eq. (\ref{sb}) is also significantly modified. 

\section{Discussion and conclusions}
\label{sec5}

In the present paper, we numerically investigated the location, stability, and basins of convergence of the equilibrium points, in the equilateral triangle configuration of the four-body problem with a radiating body. Specifically, we showed how the radiation parameter influences the dynamics of the system, in three different combinations of mass for the primaries: equal masses, two equal masses, and three different masses.

Following the tendency in this field, we used a multivariate Newton-Raphson method in order to calculate the corresponding basins of convergence. Such basins give a global view of the set of initial conditions that, after an iterative process, show a tendency to a certain fixed point. In each of the considered cases, we examine the influence of the radiation parameter on the Newton-Raphson basins of convergence. Aiming to quantify the uncertainty (complexity) of the basins, we calculated the basin entropy, monitoring its variation with the radiation parameter.

The main conclusions of this work can be summarized as follows:

\begin{enumerate}
  \item In the three considered cases, the total number of libration points decreases as $\beta\rightarrow 1$.
  \item The stability analysis suggests that, in the case of equal masses and for $\beta \in [0,1]$, the equilibrium points of the system are always linearly unstable. In the rest of the cases, the libration points are unstable if $\beta>0.01$.
  \item Taking into account the radiation factor for the solar system, we conclude that the radiation pressure and drag forces should be able to destroy the stability of the fixed points in the restricted four-body problem composed by Sun, Jupiter, Trojan asteroid and a test (dust) particle.
  \item Only in case 3 (Sun-Jupiter-Trojan Asteroid - Spacecraft system), we detected the existence of a considerable amount of non-converging initial conditions after 50000 iterations. This is due to the fact that there exists only one effective fixed point that can be reached after the iterative process. 
  \item The lowest value of the basin entropy was found in the case of two equal masses near $\beta = 0.825$, while on the other hand the highest value of $S_b$ was measured in the same case for $\beta = 1$. 
  \item It is found that in the cases of two or three equal masses, the unpredictability associated with the basin of convergence is larger for the extreme radiating case in comparison with the non-radiating one.
\end{enumerate}

\section*{Acknowledgments}
\footnotesize

FLD acknowledge financial support from Universidad de los Llanos, Grant No. CDP 2478. FLD and GAG gratefully acknowledge the financial support provided by COLCIENCIAS (Colombia), Grants Nos. 8840 and 8863.
%The authors would like to express their warmest thanks to the anonymous referee for all the apt suggestions and comments which improved both the quality as well as the clarity of the paper

\end{document}